\documentclass[compsoc, journal]{IEEEtran}
\IEEEoverridecommandlockouts

\usepackage{cite}
\usepackage{amsmath,amssymb,amsfonts}
\usepackage{graphicx}
\usepackage{textcomp}
\usepackage{xcolor}

\usepackage[utf8]{inputenc} 
\usepackage[T1]{fontenc}    
\usepackage{hyperref}     
\usepackage{url}            
\makeatletter
\g@addto@macro{\UrlBreaks}{\UrlOrds}
\makeatother

\usepackage{booktabs}       
\usepackage{nicefrac}       
\usepackage{microtype}      
\usepackage{cleveref}       
\usepackage{svg}

\usepackage[linesnumbered,ruled,noend]{algorithm2e}
\usepackage{algorithmicx}
\usepackage[section]{placeins}

\usepackage{makecell}

\usepackage{multirow}
\usepackage{bigstrut}
\usepackage{bigstrut}
\usepackage{subfig}
\usepackage{circledsteps}
\usepackage{multirow} 
\usepackage{multicol} 
\usepackage{arydshln}
\usepackage{enumitem}
\usepackage{float}

\def\BibTeX{{\rm B\kern-.05em{\sc i\kern-.025em b}\kern-.08em
    T\kern-.1667em\lower.7ex\hbox{E}\kern-.125emX}}

\author{}

\newcommand{\TheName}{\textbf{MoESys}}

\begin{document}

\title{\huge \TheName{}: A Distributed and Efficient Mixture-of-Experts Training and Inference System for Internet Services}


\author{Dianhai Yu$^*$, Liang Shen$^*$, Hongxiang Hao, Weibao Gong, Huachao Wu, Jiang Bian,~\emph{Member, IEEE},\\ Lirong Dai,~\emph{Member, IEEE}, Haoyi Xiong,~\emph{Senior Member, IEEE}

\thanks{This work was supported in part by (1) project CEIEC-2022-ZM02-0247 and (2)
Beijing Municipal Science and Technology Project (No. Z231100010323002)}

\thanks{$^*$The first two author contributed equally to this work. D. Yu, L. Shen, H. Hao, W. Gong, H. Wu, J. Bian, H. Xiong are with Baidu, Inc., Beijing, China. L. Dai is with Department of Electronic Engineering and Information Science, University of Science and Technology of China, Heifei, China. Corresponding author is Jiang Bian (email: jiangbian03@gmail.com).}}

\IEEEtitleabstractindextext{%

\begin{abstract}
  While modern internet services, such as chatbots, search engines, and online advertising, demand the use of large-scale deep neural networks (DNNs), distributed training and inference over heterogeneous computing systems are desired to facilitate these DNN models. 
  Mixture-of-Experts (MoE) is one the most common strategies to lower the cost of training subject to the overall size of models/data through gating and parallelism in a divide-and-conquer fashion. While DeepSpeed~\cite{rasley2020deepspeed} has made efforts in carrying out large-scale MoE training over heterogeneous infrastructures, the efficiency of training and inference could be further improved from several system aspects, including load balancing, communication/computation efficiency, and memory footprint limits. In this work, we present a novel \TheName{} that boosts efficiency in both large-scale training and inference. Specifically, in the training procedure, the proposed \TheName{} adopts an \emph{Elastic MoE training} strategy with \emph{2D prefetch} and \emph{Fusion communication} over \emph{Hierarchical storage}, so as to enjoy efficient parallelisms. For scalable inference in a single node, especially when the model size is larger than GPU memory, \TheName{} builds the CPU-GPU memory jointly into a ring of sections to load the model, and executes the computation tasks across the memory sections in a round-robin manner for efficient inference. 
  We carried out extensive experiments to evaluate \TheName{}, where \TheName{} successfully trains a Unified Feature Optimization~\cite{zhang2021federated} (UFO) model with a Sparsely-Gated Mixture-of-Experts model of 12B parameters in 8 days on 48 A100 GPU cards. The comparison against the state-of-the-art shows that \TheName{} outperformed DeepSpeed with 33\% higher throughput (tokens per second) in training and 13\% higher throughput in inference in general. Particularly, under unbalanced MoE Tasks, e.g., UFO, \TheName{} achieved 64\% higher throughput with 18\% lower memory footprints. 

\end{abstract}

}

\maketitle

\section{Introduction}
\label{sec:intro}

In recent years, there has been significant evolution in internet services, and the integration of artificial intelligence has made deep learning models indispensable in the internet ecosystem~\cite{al2022ai,jiang2023feynman,liu2021jizhi,bian2022afcs}. Particularly, large deep neural network (DNN) models such as BERT and GPT have gained increasing popularity due to their remarkable performance in text and language processing applications~\cite{devlin2018bert}, leading to the reliance of various internet services, including chat bots, online advertising platforms, recommender systems, search engines, and translation tools, on these models to provide users with the desired accuracy and customization~\cite{bian2023feynman,cui2020personalized,zhang2021chase,zou2021pre,liu2019enabling}.
While the utilization of large models has significantly enhanced the performance of internet services, it has come at the cost of expanding the parameter scale to tens of billions, such as the GPT-3 model with 175B parameters~\cite{brown2020language, narayanan2021efficient}, Ernie3.0 Titan with 260B parameters~\cite{wang2021ernie}, and Megatron-Turing NLG with 530B parameters~\cite{smith2022using}. However, these densely activated models necessitate abundant computing resources and extensive training time. For instance, the training of Megatron-Turing NLG with 530B parameters, one of the largest densely activated models, required three months using over 2000 NVIDIA A100 GPUs~\cite{smith2022using}, making it financially expensive and hindering the development of models with even larger parameter scales. Moreover, the inference performance of these super large-scale models seldom meets the current industrial demands~\cite{liu2021jizhi,wang2023secgnn}.

Various ad-hoc strategies have been employed to improve the efficiency of training large-scale models. One such approach is the AIBox concept, applied specifically in the training of Click-Through Rate (CTR) prediction models to reduce costs. This method involves sparsifying feature embeddings and leveraging a distributed multi-GPU setup to over-parameterize the model \cite{zhao2020distributed}. AIBox primarily focuses on certain layers for processing high-dimensional data, with an aim to scale the model. Conversely, in the realm of pre-trained language models, multi-task learning has been adopted, especially for multilingual neural machine translation \cite{caruana1997multitask}. Models like MT5 \cite{xue-etal-2021-mt5}, MASSively \cite{aharoni-etal-2019-massively}, and MultiNLI \cite{wang-etal-2020-multi} differ from densely activated models but require significant computational resources to surpass existing benchmarks.

To address these challenges, Mixture-of-Experts (MoE) based sparsely activated neural networks have been introduced for training larger models with minimal or no additional computational resources, while still achieving improved training outcomes \cite{shazeer2017outrageously, lepikhin2020gshard, fedus2021switch, shah2023novel}. MoE architectures activate only a subset of parameters based on the input data, unlike densely activated models. This selective activation results in a sub-linear increase in computational costs relative to model size. For instance, GLaM's largest variant \cite{du2021glam} possesses 1.2T parameters with 64 experts per MoE layer, yet only activates a 95B-parameter subnet (8\% of 1.2T) for each input token. Training this model saves two-thirds of the power required for GPT-3 (175B) \cite{brown2020language}, while halving the computational resources needed during inference. Despite all the benefits, MoE models still face numerous challenges and limitations, especially in computation, communication, and storage:

\begin{itemize}[leftmargin=*]
\item \emph{{Computation --}}
The computation cost per GPU remains constant in MoE models, but increases with the total number of experts. Training performance suffers due to expert imbalance, where some are overtrained and others underutilized~\cite{fedus2021switch}. Solutions include auxiliary losses~\cite{lepikhin2020gshard}, random expert selection~\cite{zuo2021taming}, and noise in routing~\cite{fedus2021switch}. However, these focus more on scheduling than computation and require substantial CPU resources. Inefficient computational task allocation and redundant operations, like H2D and D2H transfers, reduce efficiency and increase latency~\cite{he2021fastmoe}.

\item \emph{{Communication --}} 
In MoE models, imbalances in routing strategies persist despite advanced learning methods~\cite{yazdinejad2020energy,lepikhin2020gshard,fedus2021switch, yang2021m6}. Unbalanced data leads to inconsistent progress and redundant waiting in multi-task training. For example, the Switch Transformer model requires four AlltoAll communications per MoE layer, leading to performance degradation due to routing conflicts and blocking in unknown network topologies~\cite{fedus2021switch}.

\item \emph{{Storage --}} 
The memory and storage capacity limits MoE model sizes. While dense models are constrained by training time, MoE models scale better due to their sub-linear computing cost increase. A dense model with 1 trillion parameters requires 3 months to train on 3072 NVIDIA A100 GPUs, but an MoE model can be trained in weeks~\cite{narayanan2021efficient}. However, the model's scalability depends on device memory capacity. The differences in I/O latency between HBM in GPUs, CPU memory, and SSDs cause delays, necessitating efficient storage management for sparsely activated training~\cite{keckler2011gpus}.

\end{itemize}

\noindent{\textbf{Our Contributions.}} To overcome the aforementioned challenges and limitations of MoE, we introduce a novel unified framework \TheName{}, based on an open-source platform for MoE training and inference. The non-trivial contributions in \TheName{} are as follows,

\begin{itemize}[leftmargin=*]
    \item A novel distributed framework named \TheName{} is designed, which is capable of scaling MoE models to trillions of parameters, fully utilizes the clusters including HBM, CPU memory and even SSDs to break the memory wall and achieves efficient training scheduling. Notably, \TheName{} incorporates advanced techniques such as 2D prefetch scheduling and fusion communication, further enhancing the efficiency of heterogeneous storage systems.
    
    \item A new inference method based on the ring memory is employed by dynamic graph scheduling, which can integrate the computation and communication as much as possible and accelerate the inference procedure without using additional machines for larger-scale MoE models.

    \item Several effective training strategies have been initially devised in \TheName{} for NLP and CV tasks, aimed at scaling up multi-task learning without requiring additional memory. These strategies include load balancing, embedding partition, and resource-aware communication.
    
    \item We conduct comprehensive industrial-level experiments to showcase the significant performance gain using \TheName{}, where the practice in this work could benefit the future development of large-scale MoE training and inference.    

\end{itemize}

We organize the rest of this manuscript as follows. In Section 2, we review the previous efforts on the design of MoE. Section 3 introduces the novel design of \TheName{} respectively. Additionally, we reveal details of the practical implementation strategies adopted in \TheName{} in Section 4. To demonstrate the effectiveness and efficiency of \TheName{}, we conduct comprehensive experiments and analyze the results in Section 5. Finally, we conclude this work and look forward to the future direction in Section 6.

\section{Related Work}
\label{sec:related}
In this section, we review the relevant works in the field from the perspectives of large models for internet services and their training and inference systems.

\subsection{Internet Services and Large Models}

Large Language Models (LLMs) are revolutionizing internet services such as search engines, chatbots, online advertising, and cloud applications~\cite{liu2019decentralized, bian2023feynman,cui2020personalized,zhang2021chase,zou2021pre,liu2019enabling}. Organizations are increasingly using custom LLMs tailored to specific needs. These domain-specific models enhance internet service quality and customer experience, being more efficient and faster than general-purpose LLMs, particularly for applications involving proprietary data. An example is BloombergGPT~\cite{wu2023bloomberggpt}, a custom LLM by Bloomberg, which significantly impacts online finance services by rapidly evaluating financial data for risk assessments, financial sentiment analysis, and potentially automating accounting and auditing. Despite its large size of 50 billion parameters, BloombergGPT avoids traditional single-model training, favoring a Mixture-of-Experts (MoE) system for better efficiency and effectiveness. MoE models have shown great promise in natural language processing, with strategies focusing on routing enhancements~\cite{zuo2021taming, lewis2021base} to improve model quality and performance. Notice that, the GLaM~\cite{du2021glam} framework demonstrates that the largest MoE with 1.2 trillion parameters is more energy-efficient, using only one-third of the energy required for training GPT-3.

In light of the scaling law, there's a growing trend to increase model sizes. MoE-based models with billions or even trillions of parameters, like CPM-2~\cite{zhang2021cpm}, M6-T~\cite{yang2021m6}, M6-10T~\cite{lin2021m6}, and GLaM~\cite{du2021glam}, are showing superior generalization in language processing and multi-modal tasks. Baidu's UFO~\cite{zhang2021federated} model, another MoE-based framework, emphasizes deployment efficiency and big data utilization. It features a super network comprising multiple subtasks, with a routing strategy selecting the appropriate subtask for training.

\subsection{{MoE Training and Inference Systems}}

The rising popularity of the Mixture of Experts (MoE) training approach has led to the release of several open-source MoE training frameworks and systems by various scientific research bodies and corporations. DeepSpeed-MoE integrates multiple distributed parallel techniques like data parallelism and tensor slicing to effectively utilize MoE parallelism, allowing for the training of larger models. It also introduces PR-MoE, a new sparsely activated model for MoE inference, and employs model compression to reduce model sizes, alongside an efficient communication strategy to improve latency~\cite{kim2021scalable, rajbhandari2022deepspeed}. FastMoE, another distributed MoE training system, offers a user-friendly hierarchical interface and straightforward guidelines for integrating Megatron-LM and Transformer-XL with data and tensor slicing parallelism~\cite{he2021fastmoe, shoeybi2019megatron, dai2019transformer}. Unlike DeepSpeed, FastMoE focuses on reducing network traffic through an advanced optimization method. The INFMoE inference system suggests an optimal computation sequence and parameter offloading using a greedy algorithm to address workload imbalances and minimize the impact of data movement, especially when offloading to CPUs, while maintaining computational efficiency~\cite{zhang2021cpm}. Fairseq-MoE is a framework tailored for training custom models in areas like summarization, translation, and language modeling. Tutel enhances Fairseq's communication and computation capabilities, leading to a performance boost of around 40\%. Notably, these improvements in Tutel have been incorporated into DeepSpeed for MoE model training~\cite{ott2019fairseq, artetxe2021efficient, tutel}.

Furthermore, model scale and data size are two crucial factors that significantly impact the performance and effectiveness of model training. However, exploring further in this field poses a substantial challenge for scientific institutions and enterprises due to the enormous computational and storage resource requirements involved. To address this challenge, the design of sparsely activated model has emerged in recent years and gained traction in the industry. Unlike densely activated models that involve computing all parameters, the sparsely activated model dynamically selects a subset of parameters for training based on the input data. This approach enables linear parameter scaling without increasing the computational workload, thus making larger models built on the Mixture-of-Experts (MoE) architecture more feasible and efficient.

\section{\TheName{} Design}


\TheName{} is an innovative system for distributed training and inference, utilizing a Mixture-of-Experts architecture to enhance scalability and efficiency. Its main objective is to adhere to predefined memory latency goals while operating within existing storage limits. A notable advancement in this area is DeepSpeed's Zero-infinity approach \cite{rajbhandari2021zero}, which has successfully trained a model with over 30 trillion parameters using 512 V100 GPUs across NVIDIA DGX-2 nodes. This pioneering technique circumvents memory bottlenecks by fully exploiting a range of storage mediums, such as High Bandwidth Memory (HBM) in GPUs, CPU memory, and SSDs. This enables the training of exceptionally large models on singular devices. To refine storage use and boost training efficacy, both the Zero strategy \cite{rajbhandari2021zero} and a parameter prefetching method are implemented. However, there is a need to consider the reduced longevity and diminished performance of SSDs when near maximum capacity \cite{ssd}. Moreover, DeepSpeed's current prefetching approach does not accommodate the heterogeneity of parameters specific to the Mixture-of-Experts design. \TheName{} addresses these issues by introducing an innovative prefetching scheduling technique. This method enhances both training and inference by tailoring to the distinct attributes of various parameters, effectively leveraging multi-tiered storage solutions to optimize system performance.

\subsection{Overall Design of Architecture}

\begin{figure}[H]
\centering
    \includegraphics[width=0.48\textwidth]{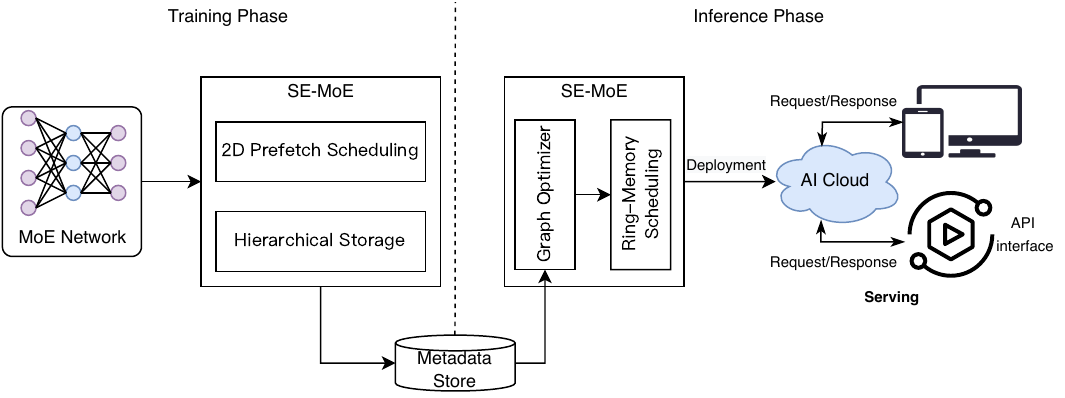}
    \vspace{-5pt}
    \caption{\TheName{}'s architecture diagram}
    \label{fig:architecture}
\end{figure}

\TheName{} employs a two-phase approach, namely the training phase and the inference phase, as illustrated in Figure~\ref{fig:architecture}. During the training phase, large-scale models are trained offline utilizing a variety of strategies. Once the model convergence is achieved, the parameters are saved for future use. On the other hand, the inference phase involves deploying the trained model to the cloud through graph optimization and pruning operations. This deployment facilitates convenient query services for users.

\subsection{Training Phase}
\label{sec:training_phase}

\begin{figure*}
    \centering 
    \includegraphics[width=0.9\textwidth]{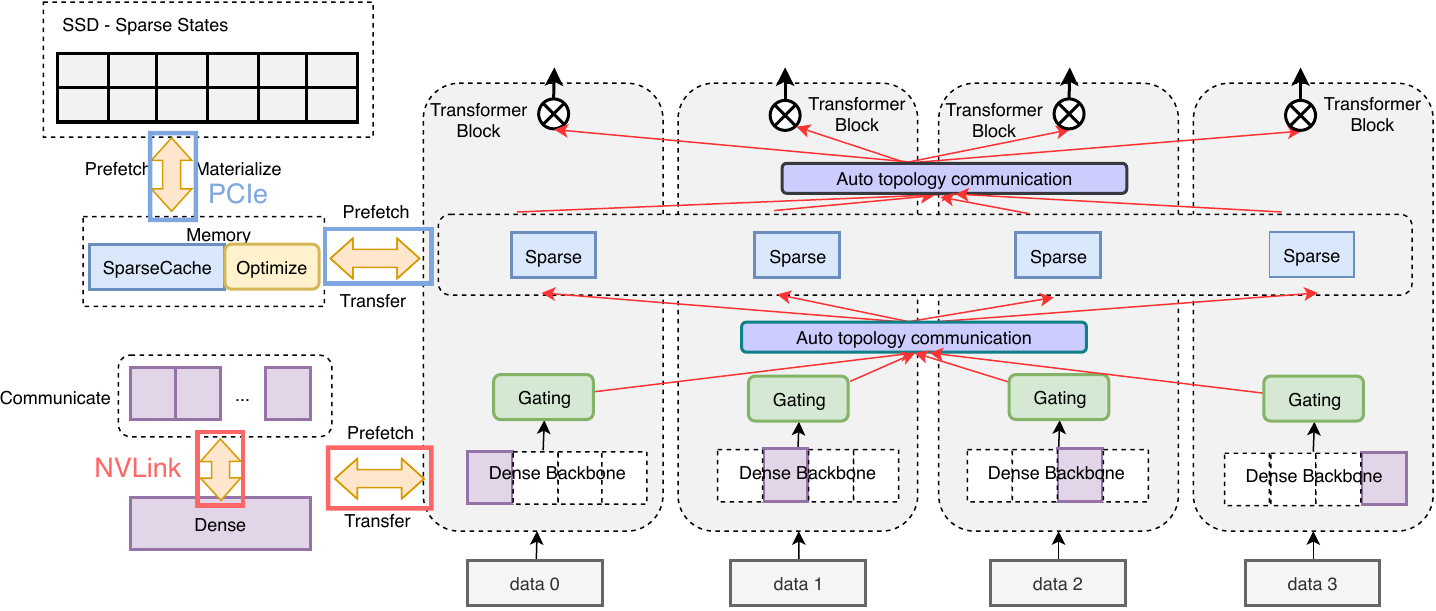}
    \vspace{-5pt}
    \caption{Overall MoE training: This is an example of the MoE training with four devices. 
    In accordance with the parameter state property of the MoE model, the parameter states are stored in both GPUs and SSDs. With this heterogeneous storage setup, we can effectively utilize the NVLink and PCIe bandwidth concurrently, leveraging their capabilities in two dimensions.}
    \label{fig:training_overall}
\end{figure*}

To enhance the efficiency of MoE training and address issues pertaining to Solid-State Drives (SSDs) and scheduling in the context of training large-scale models, a novel approach has been introduced \cite{tang2020survey}. In this method, MoE model parameters are divided into two categories according to their activation characteristics. Parameters in the first category are sparsely activated during training, such as those in the switching feed-forward network (FFN) layer, while the second category includes densely activated parameters, like those in the multi-head attention layer. Given that sparse parameters, which form a substantial part of the MoE model, may surpass GPU storage capacities, \TheName{} has restructured the MoE training system architecture, as shown in Figure~\ref{fig:training_overall}. This restructure utilizes a variety of storage mediums to meet the memory demands of both sparse and dense parameters. To counteract the performance issues arising from data transfer across different storage types, a new technique termed 2D prefetch scheduling has been implemented. The following sections will delve into a comprehensive discussion of our training framework, concentrating specifically on two principal components: \emph{Hierarchical Storage} and \emph{2D Prefetch Scheduling}.

\subsubsection{Hierarchical Storage}

In the context of large-scale Mixture-of-Experts (MoE) models, the increasing scale of parameters has led to storage becoming a significant bottleneck in model training. Typically, the stored parameter states consist of three components: trainable parameters, parameter gradients, and corresponding optimizer states. Considering the different storage media available, the storage devices can be classified into three categories: GPU-Node, CPU-Node, and SSD-Node. Since dense parameters are extensively utilized for computation and do not occupy the majority of storage space, their parameter states are stored exclusively on the GPU-Node to minimize data movement. In contrast, sparse parameters, which are selectively activated during training and consume a significant amount of storage space compared to dense parameters, have their parameter states stored on the SSD-Node and are transferred to the GPU-Node when required for calculations. By strategically allocating the corresponding parameter states to hierarchical storage based on the computational and storage characteristics of parameters, the storage capacity of devices can be maximally utilized.


In light of the constraints posed by storage nodes, this work introduces a set of theoretical formulas to articulate the correlation between various storage devices and the storage requirements of parameter states when utilizing the ADAM optimizer \cite{kingma2014adam}. Typically, each storage device is configured with eight GPUs. We denote the aggregate count of dense and sparse parameters as $D$ and $S$ respectively, and $L$ as the total number of MoE layers. The capacities of SSD memory, CPU memory, and GPU memory in a single device are represented by $M_{SSD}$, $M_{CPU}$, and $M_{GPU}$, in that order. Moreover, $N$ signifies the quantity of devices. We also introduce a variable, $\alpha$, to quantify the likelihood of activation of sparse parameters during training, with $\alpha$ ranging between 0 and 1.

For the GPU-Node, it stores the dense parameter states used in forward propagation (\hypertarget{A}{\hyperlink{B}{FWD}}), backward propagation (\hypertarget{C}{\hyperlink{D}{BWD}}), and parameter updating. This includes parameters such as param fp16, grad fp16, master param fp32, momentum fp32, variance fp32, with a total size of $2D+2D+4D+4D+4D=16D$ bytes. Furthermore, it accommodates sparse parameters and their corresponding gradients, with a size of $4\alpha S/L$ bytes, accounting for the selective activation of sparse parameters. The CPU-Node serves as a cache to hold high-frequency sparse parameter states, occupying $16\alpha S$ bytes. Lastly, the SSD-Node stores all sparse parameter states on the device, including master param fp32, momentum fp32, and variance fp32, with a size of $12S$ bytes.

\begin{equation}
\begin{aligned}
    &\textbf{GPU-Node}:&  16D + 4\alpha S/L &\leq M_{GPU} \cdot N\\
  &\textbf{CPU-Node}:&  16\alpha S &\leq M_{CPU} \cdot N \\
    &\textbf{SSD-Node}:&  12S &\leq M_{SSD} \cdot N
\end{aligned}
\label{equ:Node storage}
\end{equation}

The scale of the entire MoE model:
\begin{equation}
\begin{aligned}
    P = S + D 
\end{aligned}
\label{equ:parameters}
\end{equation}


The storage mechanism for sparse parameters typically involves saving them on SSDs. Nonetheless, SSDs encounter limitations due to their flash media, limited PCIe bandwidth, and constraints of the NVMe protocol. These factors contribute to increased latency and a restricted number of erasures, posing challenges in MoE training scenarios that require frequent write operations. To address these challenges, we turn our focus to Intel Optane Persistent Memory (Optane PMem) \cite{Optane}, an innovative storage medium that merges the benefits of byte-level addressing, similar to DRAM, with the long-term storage ability of SSDs. Optane PMem is connected to the CPU's integrated memory controller (IMC) via the DIMM (Dual Inline Memory Module) interface and communicates using DDR-T, a protocol developed for DDR4's electrical/mechanical interface. This configuration allows for byte-level addressing through CPU commands, enhancing bandwidth and decreasing latency. Significantly, Optane PMem functions in two modes: memory mode and AppDirect mode. For our specific requirement of storing parameter files on Optane PMem, we choose the AppDirect mode and set the namespace to FSDAX. By exploiting the features of Ext4, direct load and store operations are possible, circumventing both the CPU's page cache and the kernel, which facilitates seamless data transfer free from interruptions or context switches.

\subsubsection{2D Prefetch Scheduling}



The implementation of hierarchical storage for the preservation of both sparse and dense parameter states in MoE training introduces considerable time overhead due to the necessity of transferring these states across various devices. To mitigate this, a 2D prefetch scheduling strategy is proposed, allowing for the simultaneous processing of dense and sparse schedules during MoE training. This strategy facilitates the concurrent computation of parameters with the scheduling procedure.

In greater detail, this strategy, particularly when applied to the dense parameter subset as defined by the ZeRO-3 strategy, enables prefetching of the entire dense parameter set post inter-rank communication along the horizontal axis, utilizing the rapid transfer speeds of NVLink. This approach is instrumental in achieving data parallelism, as demonstrated in Algorithm~\ref{alg:dense Scheduling}. In this methodology, prefetching occurs alongside the computation and communication processes of the current layer. To be more specific, while the \(i^{th}\) layer undergoes computation and communication, prefetch scheduling for the \((i+1)^{th}\) layer's parameters is conducted in parallel. This simultaneous prefetching approach guarantees the readiness of parameters for the subsequent layer when required, significantly reducing idle times and boosting overall computational efficiency.

\IncMargin{1.0em}
\begin{algorithm}[ht]
  \caption{Scheduling on Dense Parameters}
    \label{alg:dense Scheduling}
    \SetAlgoLined
    ${d_{i}^{'}}$: Dense parameter state slices in $i^{th}$ layer\\ 
    $d_{i}$: total dense parameters in $i^{th}$ layer \\

  \SetInd{0.61em}{0.61em}
  \BlankLine
  
  \SetKwFunction{FDesne}{DenseSchedule}
    \SetKwProg{Fn}{Function}{:}{}
    \Fn{\FDesne{i}}{
        Get dense parameters in $i^{th}$ layer $d_{slice}$ \\
        $d$ = $AllGather({d_{i}^{'}}$)
    }
    \textbf{End Function}
    
\end{algorithm}
\DecMargin{1.0em}

In a similar vein, the prefetching of sparse parameters takes place through the PCIe bandwidth in the vertical dimension of the device. Given that sparse parameters are stored in SSDs, we mitigate access to SSDs for sparse parameter states by implementing a cache mechanism in the CPU memory, akin to the LFU (Least Frequently Used) mechanism~\cite{sokolinsky2004lfu}. CPU caches are responsible for storing selectively activated sparse parameter states used in FWD/BWD calculations and parameter updates. When a prefetch request is received, it is prioritized to retrieve the requested sparse parameters from the CPU caches. If these parameters are not found in the CPU caches, they are subsequently retrieved from the SSDs. Moreover, when the CPU caches become full or when the sparse parameter update cycle period is reached, the sparse parameter states from the CPU caches are used to update the corresponding parameter states on the SSDs.

\IncMargin{1.0em}
\begin{algorithm}[ht]
  \caption{Scheduling on Sparse Parameters}
    \label{alg:Sparse Scheduling}
    \SetAlgoLined
  \SetKwProg{Fn}{Function}{}{end}
    \textbf{Parameters:}\\
    $p_{s}$: {\footnotesize sparse parameter states in $i^{th}$ layer} \\ 
    $caches_{cpu}$: {\footnotesize CPU caches} \\
    $CPU_{size}$: {\footnotesize the maximum capacity of the CPU caches to store sparse parameter states}\\
    $hits$: {\footnotesize the frequency of hits for a specific sparse parameter in the hash table} \\
    $threshold$: {\footnotesize hit threshold} \\
    $\beta $: {\footnotesize attenuation coefficient}\\
    $K$: {\footnotesize the step size of moving average}\\
    $steps=0$: {\footnotesize cycle steps}\\
    $acc_{caches}=0$: {\footnotesize cumulative caches} \\

  \SetInd{0.61em}{0.61em}
  \BlankLine

    \BlankLine
    \SetKwFunction{FSparse}{SparseSchedule}
    \SetKwProg{Fn}{Function}{:}{}
    \Fn{\FSparse{i}}{
        \uIf{$p_{s}$ in $caches_{cpu}$}{Get $p_{s}$ from $caches_{cpu}$ \\ $hits[p_{s}]$ += 1} 
        \ElseIf{$acc_{caches}$ + 1 < $CPU_{size}$}{$hits[p_{s}]$ = 1 \\ $acc_{caches}$ += 1 \\ Fetch $p_{s}$ from SSDs to $caches_{cpu}$}
        \Else{
            \ForEach{$p_{a}$ in $hits$}{
                $hit_{a} = hits[p_{a}]$ \\
                \uIf{$hit_{a} \geq threshold$ and $\min(hits.values()) == hit_{a}$}{
                Update the states of $p_{a}$ on SSDs  \\
                Delete the states of $p_{a}$ in $caches_{cpu}$ \\
                Delete $hits[p_{a}]$ \\
                Fetch $p_{s}$ from SSDs to $caches_{cpu}$}}
    }
    $steps$ += 1 \\
    \uIf{$steps$ == $K$}{
        $hits$ $\cdot$ $\beta$  \algorithmiccomment{{\scriptsize moving average}}\\
        $steps = 0$ 
    }
    $p_{s} \longrightarrow GPU$ \algorithmiccomment{{\scriptsize transfer $p_{s}$ to the corresponding GPU}} \\
    }
    \textbf{End Function}

\end{algorithm}
\DecMargin{1.0em}

As the CPU memory on each machine only caches frequently activated sparse parameters, we only need to prefetch the parameters of one or more expert layers, which are cached in the CPU memory, to the corresponding GPU memory in advance. By prefetching parameters in advance, the waiting time for computation can be significantly reduced. From a global perspective, by utilizing the bandwidth of NVLink and PCIe in two dimensions, we can simultaneously prefetch dense and sparse parameters, effectively reducing the scheduling gap caused by heterogeneous storage and greatly enhancing training efficiency. In the following sections, we present a detailed explanation of the CPU cache mechanism, as depicted in Algorithm~\ref{alg:Sparse Scheduling}. Additionally, we maintain historical hit information for each sparse parameter, which is recorded in a hash table referred to as $hits$. Specifically, if a parameter $p_s$ is requested and has been used in the previous FWD, we increment its count in the $hits$ table. When the CPU caches have reached their maximum capacity, we update the sparse parameter states with the lowest hit frequency that surpasses the hit threshold.

In the MoE model training, each node determines whether to activate its experts in the next iteration based on the recorded expert selection results and the maintained experts' information. If activation is needed, further decisions are made based on the historical hit information recorded in a hash table to determine whether to send prefetch requests. Firstly, to avoid introducing additional CPU operations before sending prefetch requests, it is essential to place the hash table that records historical hit information on the GPU Node. Since each node only stores a portion of the sparse parameters in the SSD (not the full set), it is only necessary to maintain historical hit information for the corresponding sparse parameters. This approach distributes the GPU space cost across all computing nodes, making it negligible. Secondly, the process of selecting experts by the Gate network inherently requires All-to-All communication to synchronize the selection results across each node in the Expert Parallelism Group. The prefetch scheduling simply reuses the results of this All-to-All communication, so no additional communication operations are introduced. Additionally, the time complexity of a hash table is O(1), meaning each prefetch operation involving searches, insertions, or deletions can be completed in constant time, thus not introducing additional computational costs.


The distinct and non-interfering characteristics of dense and sparse parameters in the model facilitate the simultaneous implementation of prefetch strategies. This approach optimally leverages the bandwidth capacities of both NVLink and PCIe. While the GPU is engaged in prefetching the parameter state for the upcoming layer, it can also simultaneously execute computations for the current layer. This dual-operation mode efficiently combines the tasks of computation and parameter readiness.

\subsection{Inference Phase}

\label{sec:inference_phase}


Numerous studies~\cite{du2021glam, artetxe2021efficient} have demonstrated that Mixture-of-Experts (MoE) models exhibit significantly higher training efficiency compared to dense models. However, during inference, the presence of numerous parameters, many of which are ineffective, poses a challenge of increased storage requirements compared to dense models. Knowledge distillation~\cite{fedus2021switch, huang20221+, shleifer2020pre, sanh2019distilbert} has emerged as a popular approach for reducing model size while preserving accuracy. In this context, DeepSpeed~\cite{rajbhandari2022deepspeed} has proposed the Mixture-of-Students (MoS) architecture to enhance the accuracy of the student models. Specifically, to achieve low latency and high throughput at a large scale for MoE models, various parallelism techniques have been devised~\cite{rajbhandari2022deepspeed}, including expert-slicing, expert parallelism, tensor-slicing, and others. However, the inference of MoE models at an unprecedented scale often neglects the consideration of multiple storage devices when the number of machines is limited.

In the following subsections, we present the approach adopted by \TheName{} to achieve high efficiency throughout the training and inference deployment. We optimize the graph training process and propose innovations in the MoE inference architecture based on ring memory. This architecture addresses the memory wall challenge and ensures optimal performance to the greatest extent possible.

\subsubsection{Graph Optimization}


\begin{figure}
    \includegraphics[width=0.47\textwidth]{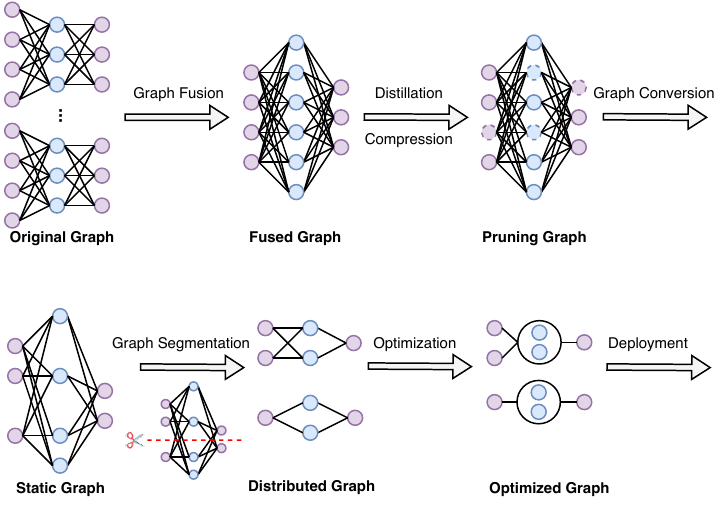}
    \vspace{-5pt}
    \caption{Inference Pipeline in MoE.}
    \label{fig:inference_overall}
\end{figure}

The training phase of \TheName{} incorporates dynamic graph training, which offers significant advantages in terms of debugging and flexibility. In contrast, for enhanced stability and efficiency, the inference and deployment stages utilize a static graph. Figure~\ref{fig:inference_overall} illustrates the overall process of inference, which comprises six key steps:
\begin{itemize}[leftmargin=*]
    \item Graph Fusion -- The original graph is merged with the corresponding distributed strategy to accommodate ultra-large-scale distributed training. This step involves eliminating parameter redundancy.

    \item Distillation and Compression -- The numerous experts in the teacher network are compressed through distillation and compression techniques, resulting in a student network with fewer experts.

    \item Graph Conversion -- The dynamic graph is converted into a static graph to enable subsequent optimization and deployment processes. Due to space limit, we introduce the detailed strategy of conversion in the external link\footnote{\url{https://www.paddlepaddle.org.cn/documentation/docs/en/guides/jit/basic_usage_en.html}}.

    \item Graph Segmentation -- Based on available inference resources and specific requirements, a rational distributed strategy is chosen either manually or automatically to partition the static graph into multiple distributed sub-graphs. Additional communication is added as needed.

    \item Optimization -- Pertinent Intermediate Representation (IR) Pass optimizations, such as kernel fusion, are applied to the distributed sub-graphs to further improve inference performance.

    \item Deployment -- The optimized sub-graphs are deployed on servers to provide efficient and reliable services.
\end{itemize}

It is important to note that \TheName{} combines highly optimized transformers and MoE-related kernels. We leverage optimized methods, such as Fused Multi-head Attention, which have been successfully employed in NVIDIA's BERT implementation for MLPerf 1.1~\cite{mattson2020mlperf}. These optimizations effectively reduce kernel launch time. For the MoE model, we have developed unique kernels to improve H2D/D2H (Host-to-Device/Device-to-Host) transfer time by utilizing CUDA Pinned Memory and customizing AlltoAll communication. Our aim is to minimize the number of layer transitions as much as possible. The details of these optimizations and their impact on the performance of \TheName{} are presented and discussed in Section~\ref{sec:infer_exp}.


\subsubsection{Ring Memory Offloading}
\label{sec:ring_memory}


In order to facilitate the inference of large-scale MoE models with limited resources, it is essential to employ an offloading strategy to address the storage challenge. However, the speed of data movement often becomes a limiting factor for inference performance. Consequently, numerous methods aim to conceal the impact of data movement by maximizing the overlap between data movement and inference calculations, thereby reducing the waiting time for calculations. In this work, we propose a dynamic scheduling strategy for offloading sparse parameters, specifically expert parameters in the MoE model. The objective is to maintain efficient performance by concurrently moving parameters from CPU memory while performing inference computations in GPU memory. By overlapping these operations, we aim to minimize the overall latency and enhance the efficiency of the inference process.

\begin{figure}[H]
    \includegraphics[width=0.4\textwidth]{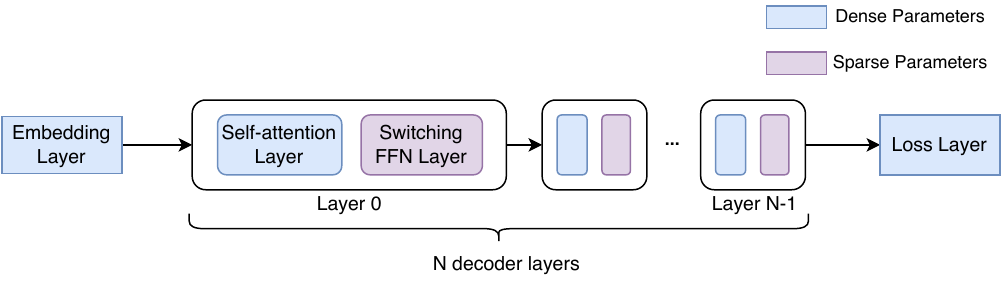}
    \vspace{-5pt}
    \caption{Switching Layers in MoE Inference Model.}
    \label{fig:infer_model}
\end{figure}

The structure of the MoE model during its inference stage, illustrated in Figure~\ref{fig:infer_model}, demonstrates a layer-specific independence of parameters, reminiscent of the switch transformer architecture \cite{fedus2021switch}. This design feature enables the staggering of computation and offloading tasks, thereby facilitating their concurrent execution. Considering an MoE inference model comprising $N$ decoder layers, each layer's expert parameters are replicated $N$ times and stored on the CPU device. Concurrently, other parameters, such as embeddings, are maintained within the dense buffer of the GPU device. In addition, $K$ replicas of the expert parameters are also cached within the GPU device.

\begin{figure}
\centering
    \includegraphics[width=0.35\textwidth]{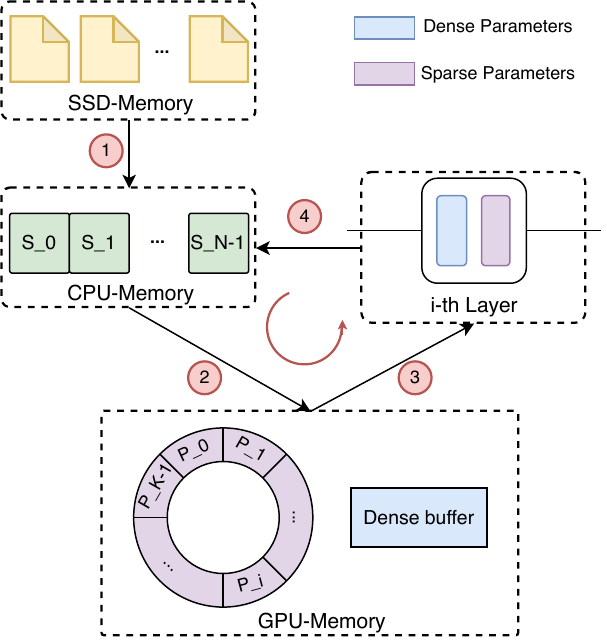}
    \vspace{-5pt}
  \caption{The scheduling and timeline of the ring memory offloading process can be summarized into the following essential steps:
    \Circled{1} Load $N$ copies of parameters from files in SSD memory, \Circled{2} Load $K$ copies of parameters from CPU memory,  
    \Circled{3} Execute the computation for the $i$-th layer, \Circled{4} Release the $i$-th parameter and trigger asynchronous copy process to replace $P_i$ with $S_{K+i}$.}
    \label{fig:infer}
\end{figure}


As depicted in Figure~\ref{fig:infer}, upon completion of the computation pertaining to the $i$-th layer, the corresponding parameter $P_i$ in the GPU memory can be released. Concurrently, the $S_{K+i}$ expert parameter of the $(K+i)$-th layer can be asynchronously loaded from the CPU memory to occupy the previously utilized space by $P_i$. This procedure, referred to as calculation-released-load, facilitates the maintenance of a fixed number of $K$ expert parameter duplicates on the GPU device. These duplicates are stored in the ring memory, thereby mitigating memory fragmentation. By leveraging distinct CUDA streams, the expert loading from the CPU and the computation process can be partially overlapped. Moreover, by ensuring a substantial ring memory size and incorporating a greater number of decoder layers in the MoE inference model, the level of overlap can be significantly optimized. For an evaluation of the inference performance using the ring memory approach, please consult Section~\ref{sec:infer_exp}.


\section{Implementation strategies}

\label{sec:eff_methods}

The distinctive architecture of the MoE model gives rise to inherent challenges in both the training and inference processes. In order to tackle the issue of load imbalance caused by uneven input data, we have devised the Elastic MoE Training approach. Furthermore, recognizing the significant involvement of cross-machine communication in MoE, we have delved into Resource-aware Communication techniques to enhance efficiency across diverse clusters. Lastly, to overcome storage limitations stemming from the use of oversized vocabularies in various tasks, we have developed and implemented a novel embedding partition method within the framework of data parallelism, distinct from the approach employed in tensor-slicing parallelism.

\subsection{Elastic MoE Training}
\label{sec:elastic_training}


Load imbalance significantly influences training efficiency, especially in multi-task training scenarios employing the MoE framework. A prominent instance of this is seen in the UFO task, where the differential input data volumes across various tasks lead to unequal computation durations, thus exacerbating load imbalances. This discrepancy manifests in two primary forms: one, the breaching of memory capacity limits due to the processing of disproportionately large batch sizes by individual task nodes, a consequence of aggregating data from other nodes; and two, the retardation of synchronous communication caused by the lag of the slowest node. This phenomenon, known as the "Cask Effect" \cite{chen2019cask}, results in reduced computational efficiency.

\begin{figure}
    \centering
    \subfloat[Original]{\includegraphics[width=0.15\textwidth]{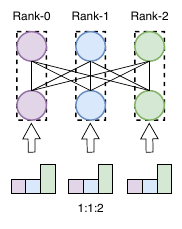}
    \label{fig:Original}}
    \hfil
    \subfloat[Scale down]{\includegraphics[width=0.14\textwidth]{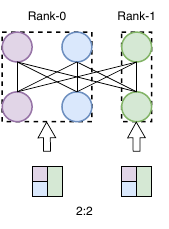}
    \label{fig:Scale down}}
    \hfil
    \vspace{-3mm}
    \subfloat[Scale up]{\includegraphics[width=0.2\textwidth]{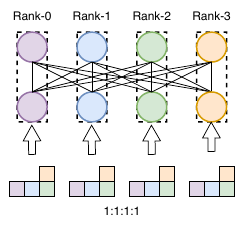}
    \label{fig:Scale up}}
    \vspace{-5pt}
    \caption{Different methods supported by elastic MoE training: (a) the original 
    training with load imbalance, in which the ratio of each node data quantity is 1:1:2;
    (b) combining multiple nodes with light-duty tasks, in which the ratio of each node data quantity is 2:2;
    (c) adding extra nodes to handle heavy-duty tasks, in which the ratio of each node data quantity is 1:1:1:1.}
    \label{fig:load imbalance}
\end{figure}

To tackle these challenges, we implement the elastic MoE training method, which dynamically adjusts the number of training nodes to ensure load balance based on the estimated workload of each task. In practical terms, for lighter tasks, combining multiple nodes proves more resource-efficient, as long as storage capacity is not a limiting factor (see Figure~\ref{fig:Scale down}). Conversely, for heavier tasks, we introduce additional nodes to distribute the workload across more computing resources. Simultaneously, we partition the input data of heavy-duty tasks to achieve load balance and employ data parallelism to ensure parameter synchronization (see Figure~\ref{fig:Scale up}). These elastic training approaches effectively mitigate performance degradation resulting from load imbalance. You can find detailed performance comparisons in Section~\ref{exp:ela_train}.

In elastic training, load imbalance often leads to situations where some devices are idle, waiting for others to complete their computations, creating what is known as "bubbles." These bubbles not only reduce computational efficiency but also increase training costs. To address this, we have introduced a method of scaling up or down the computational devices dynamically.
This method aims to reduce waiting times and enhance the FLOPS (floating-point operations per second) utilization rate per device, thereby boosting the throughput (tokens/s/card) of each compute device. The decision to scale up or down should be based on specific training requirements and cost considerations:
\begin{itemize}[leftmargin=*]
    \item Upscaling: When there is a need to increase the overall end-to-end throughput, we typically scale up by adding more compute devices. This reduces the total training time for the model.
    \item Downscaling: In situations where resources are constrained, and cost control is crucial, we opt for downscaling by reducing the number of compute devices, thereby lowering the cost of training the model.
\end{itemize}
Irrespective of the scaling direction, both methods effectively enhance the FLOPS utilization rate, reduce idle waiting times for compute devices, and decrease the overall "bubble" time during training.

\subsection{Resource-Aware Communication}

In the process of training and inference of the MoE model, a significant volume of AlltoAll communication is necessitated between devices in the context of expert parallelism. This communication procedure has the potential to become a performance bottleneck, as multiple instances of AlltoAll communication may contend for the finite network resources concurrently. Upon analysis of network topologies in typical clusters, it is observed that data interaction across clusters exhibits relatively slower transmission speeds compared to interactions within a single cluster. This disparity arises from factors such as congested message pathways and higher traffic costs associated with inter-cluster communication.

\begin{figure}
    \centering 
    \includegraphics[scale=0.42]{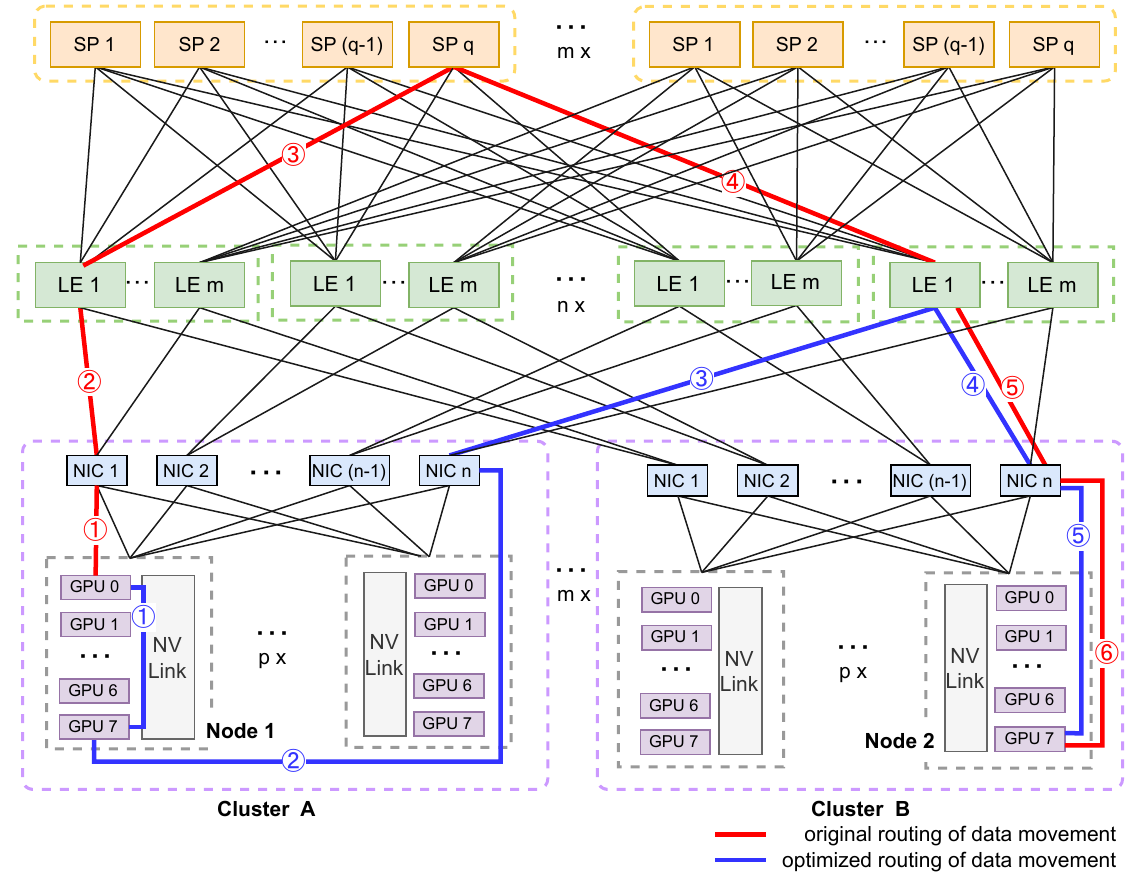}
    \vspace{-5pt}
    \caption{Network topology and message pathways for data movement}
    \label{fig:network_topology}
\end{figure}

By utilizing NVLink, intra-node communication incurs minimal time and resource overhead as it avoids traversing any Network Interface Cards (NICs) or switches. However, inter-node communication within a cluster or across clusters necessitates traversing a congested message path that involves NICs and switches~\cite{ling2012extensive}, resulting in increased time consumption for traffic scheduling. Consider a network consisting of $m$ clusters and $p$ nodes that share a common set of NICs within each cluster. The leaf switches (LE) and spin switches (SP) are organized into $n$ and $m$ groups, respectively. As depicted in Figure~\ref{fig:network_topology}, the leaf switches of the $i$-th group establish direct connections only to NICs with a rank of $i$ from different clusters. The spin switches facilitate communication between leaf switches. It is important to note that the bandwidth of the spin switch is lower than that of the leaf switch. Hence, it is preferable to maximize the utilization of the leaf switch for data exchange, aiming for improved performance. For instance, let us consider a scenario where all GPU0s are connected to NIC1 and all GPU7s are connected to NICn. We observe that data movement between GPU0 of Node1 in Cluster A and GPU7 of Node2 in Cluster B traverses the switch routing path $[LE1, SPq, LE1]$ as indicated by the red lines. This incurs higher communication costs and the potential for resource contention with other interactions. An alternative approach involves a two-step process: first, transferring data from GPU0 to GPU7 within Node1 using NVLink, and then performing cross-cluster communication between the corresponding pair of NICs with rank 7, without crossing any switches except $LE1$. This is depicted by the blue lines. Such an approach enables the optimal utilization of NVSwitch bandwidth and enhances network traffic optimization.

\begin{figure}
    \centering 
    \includegraphics[width=\columnwidth]{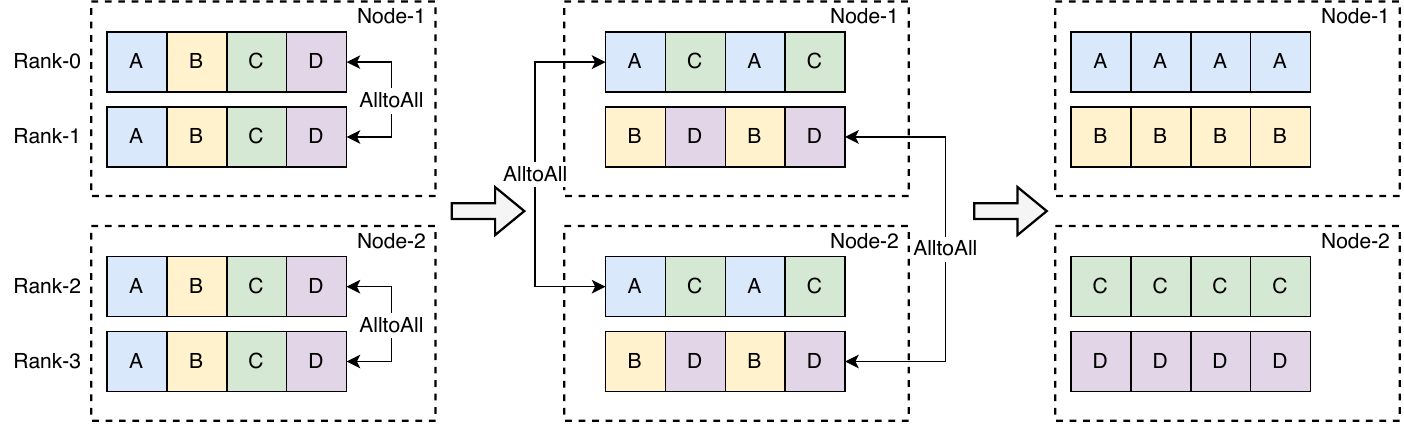}
    \vspace{-15pt}
    \caption{Hierarchical AlltoAll}
    \label{fig:hierarchical_alltoall}
\end{figure}

Hence, the speed at which data is exchanged between GPUs of the same rank within a node surpasses that of GPUs with different ranks within the same node. To leverage the network topology's characteristics, we propose an optimized approach for Hierarchical AlltoAll communication that takes into account the available resources during both training and inference. As depicted in Figure~\ref{fig:hierarchical_alltoall}, to avoid cross-node communication involving GPUs of different ranks, we initially utilize intra-node AlltoAll communication through NVSwitch connections to collect the data. Afterward, we group GPUs with identical ranks for inter-node AlltoAll communication, allowing communication across machines without incurring unnecessary costs related to crossing different channels.

Furthermore, this approach enhances peer-to-peer communication between nodes by a factor of $p$, where $p$ denotes the number of GPUs within a single node. This increase in inter-node communication capacity enables the optimal utilization of inter-node bandwidth. In contrast to DeepSpeed's AlltoAll design~\cite{rajbhandari2022deepspeed}, which is primarily aimed at addressing the issue of small per-port communication volumes in all-to-all communication through a layered approach for tensor fusion to facilitate larger packet communication, our approach is specifically tailored to the network topology of our experimental cluster. We concentrate on maximizing the utilization of NVLink connections and mitigating network congestion. Therefore, our Hierarchical AlltoAll structure is a direct response to our specific network topology and would adapt if the cluster's topology were to change. This distinction underscores the customization of our method to our particular hardware and network architecture, which differs from the more generalized approach adopted by DeepSpeed.

\subsection{Embedding Partition in Data Parallelism}

In the context of ultra-large-scale model training, the embedding table often constitutes the largest parameter set within the entire model, thereby necessitating restrictions on its storage due to the model's scale. Numerous studies have focused on researching embedding partitioning techniques. For instance, Megatron~\cite{shoeybi2019megatron} has successfully employed column-wise partitioning of the embedding table in tensor-slicing parallelism to reduce training memory requirements. Additionally, EmbRace~\cite{li2021embrace} has proposed a column-wise partitioning approach within the embedding table to achieve more balanced communication. However, an efficient processing method for handling embedding partitioning in scenarios where the input data for each process is inconsistent remains elusive. In such cases, ensuring efficient processing becomes challenging due to the varying nature of the inputs across different devices.

\begin{figure}
    \centering
    \subfloat[Forward stage]{\includegraphics[width=0.245\textwidth]{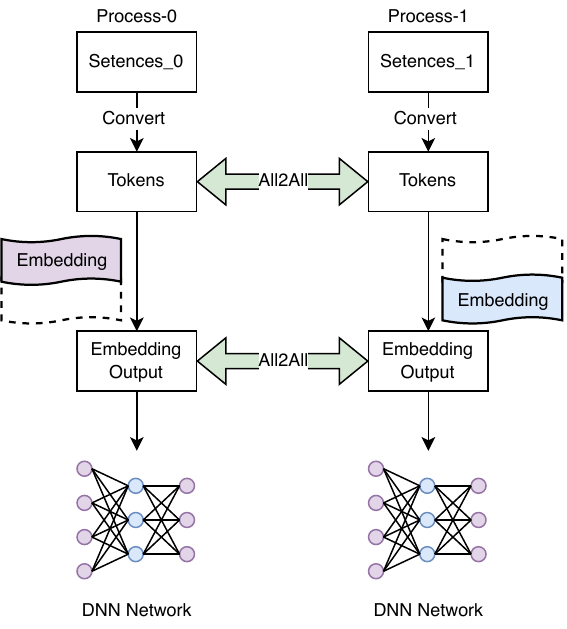}
    \label{fig:fp_dp_embedding}}
    \centering
    \subfloat[Backward and optimization stage]{\includegraphics[width=0.245\textwidth]{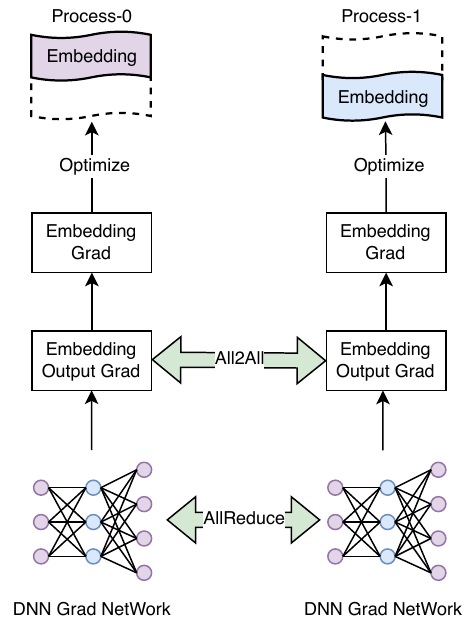}
    \label{fig:bp_dp_embedding}}
    \vspace{-5pt}
    \caption{Example data flow of Embedding Partition in data parallelism. The embedding table is column-wise partitioned among processes. In the forward stage, AlltoAll communication is called twice: one is for exchanging input data and another is for exchanging embedding lookup results. In the backward stage, 
    AlltoAll is called once to swap the gradient of the embedding table and using the gradients updates the embedding table.}
    \label{fig:dp_embedding}
\end{figure}

In this study, our primary focus is on addressing the challenge of embedding partitioning within the context of data parallelism, as illustrated in Figure~\ref{fig:dp_embedding}. To elaborate, when we consider an embedding table with dimensions $[V, H]$ being distributed among $N$ training processes, we employ a column-wise partitioning scheme. This scheme assigns a $[V, \frac{H}{N}]$ shard to each worker. Consequently, each training process possesses an embedding representation that pertains only to a subset of the vocabulary. As a result, before querying the embedding table, the input data of each process must be exchanged with other processes through AlltoAll communication. This exchange allows the acquisition of embedding results corresponding to the local partial vocabulary. Subsequently, to obtain accurate results for the input data processed by each worker, the embedding results are exchanged once more through AlltoAll communication, effectively serving as the inverse procedure of the previous communication step. It is essential to note that during the backward stage, the gradient information also needs to be exchanged to recover the embedding table gradient.

In contrast to the traditional embedding sharding approach~\cite{mudigere2022software}, which is similar to tensor model parallelism and partitions along the vocabulary dimension, our method is specifically designed for data parallelism. Given that each GPU card processes different data, sharding along the vocabulary dimension is not feasible. Instead, we employ sharding along the hidden\_size dimension of the vocabulary, ensuring that each computing device can access the complete vocabulary. Additionally, we utilize AlltoAll communication to complete the hidden layer, accommodating the varying data inputs across devices.

Significantly, this approach effectively reduces the storage requirements of the embedding table within the data parallelism framework. It achieves this by introducing only three instances of AlltoAll communication and eliminating the need for AllReduce synchronization for embedding table gradients in data parallelism.

\section{Experiment}
\label{sec:exp}
In this section, we present a comprehensive experimental evaluation of MoE models using the proposed \TheName{} system. Our evaluation focuses on both training and inference aspects of the MoE models. In the training phase, we assess the efficiency of the MoE-based GPT model across different configurations. For the inference phase, we analyze the performance of the ring memory-based offloading strategy, considering various model sizes. Moreover, we investigate several efficient methods implemented in the \TheName{} system, including the widely-used UFO model. It is important to note that our results reporting disregards the performance of the individual models themselves. This is because the converged models, whether based on baseline MoEs or utilizing \TheName{}, achieve comparable performance levels. Hence, our evaluation primarily focuses on the efficacy and efficiency of the proposed \TheName{} system.

\subsection{Platform}
MoESys is implemented based on the PaddleFleetX~\cite{paddlefleetx} architecture of PaddlePaddle. After some fundamental performance optimizations, PaddleFleetX demonstrates certain performance advantages over other standard models. For a more direct comparison, we would like to highlight the performance comparison between Paddle and Megatron as of March 2023. This comparison provides valuable insights into how PaddleFleetX, and by extension MoESys, stands in relation to other prominent frameworks in terms of efficiency and effectiveness. Please refer to Table~\ref{fig:platform} for detailed comparative data. The experimental data demonstrates PaddleFleetX's enhanced performance over Megatron-LM across several model configurations, with a notable throughput superiority that ranges from a substantial 14.2\% for smaller models (0.35 billion parameters) to a marginal 0.4\% for the extensive 175 billion parameter models. This improved throughput is consistent despite marginal differences in memory usage where Megatron-LM occasionally leads, particularly with smaller models. PaddleFleetX also showcases a more efficient utilization of GPU resources, as evidenced by higher TFLOPS/s per GPU and a closer approach to the theoretical peak FLOP/s utilization across varying model sizes. 

\begin{table*}
  \centering
  \caption{The standard model's performance on the established baselines.}
  \vspace{-3mm}
  \resizebox{0.9\textwidth}{!}{%
        \begin{tabular}{|c|c|c|c|c|c|c|c|c|c|cccc|c|}
\hline
\multirow{2}{*}{\textbf{\begin{tabular}[c]{@{}c@{}}Parameters \\ (Billions)\end{tabular}}} & \multirow{2}{*}{\textbf{\begin{tabular}[c]{@{}c@{}}Hidden\\ size\end{tabular}}} & \multirow{2}{*}{\textbf{Layers}} & \multirow{2}{*}{\textbf{\begin{tabular}[c]{@{}c@{}}Attention\\ heads\end{tabular}}} & \multirow{2}{*}{\textbf{GPUs}} & \multirow{2}{*}{\textbf{\begin{tabular}[c]{@{}c@{}}Data\\ parallel\end{tabular}}} & \multirow{2}{*}{\textbf{\begin{tabular}[c]{@{}c@{}}Group\\ sharded\\ parallel\end{tabular}}} & \multirow{2}{*}{\textbf{\begin{tabular}[c]{@{}c@{}}Tensor\\ parallel\end{tabular}}} & \multirow{2}{*}{\textbf{\begin{tabular}[c]{@{}c@{}}Pipeline\\ parallel\end{tabular}}} & \multirow{2}{*}{\textbf{\begin{tabular}[c]{@{}c@{}}Batch\\ size\end{tabular}}} & \multicolumn{4}{c|}{\textbf{\begin{tabular}[c]{@{}c@{}}PaddleFleetX \\ (vs Megatron-LM )\end{tabular}}}                                                                                                                                                                                                                                                                & \multirow{2}{*}{\textbf{\begin{tabular}[c]{@{}c@{}}Difference\\ of\\ throughput\end{tabular}}} \\ \cline{11-14}
                                                                                           &                                                                                 &                                  &                                                                                     &                                &                                                                                   &                                                                                              &                                                                                     &                                                                                       &                                                                                & \multicolumn{1}{c|}{\textbf{\begin{tabular}[c]{@{}c@{}}Throughput\\ (tokens/s)\end{tabular}}} & \multicolumn{1}{c|}{\textbf{\begin{tabular}[c]{@{}c@{}}Memory\\ usage (MB)\end{tabular}}} & \multicolumn{1}{c|}{\textbf{\begin{tabular}[c]{@{}c@{}}TFLOPS/s\\ per GPU\end{tabular}}} & \textbf{\begin{tabular}[c]{@{}c@{}}Theoratical\\ peak FLOP/s (\%)\end{tabular}} &                                                                                                \\ \hline
0.35                                                                                       & 1024                                                                            & 24                               & 16                                                                                  & 8                              & 8                                                                                 & 1                                                                                            & 1                                                                                   & 1                                                                                     & 64                                                                             & \multicolumn{1}{c|}{\begin{tabular}[c]{@{}c@{}}291,585\\ (255,401)\end{tabular}}              & \multicolumn{1}{c|}{\begin{tabular}[c]{@{}c@{}}31,919\\ (33,217)\end{tabular}}            & \multicolumn{1}{c|}{\begin{tabular}[c]{@{}c@{}}102\\ (89)\end{tabular}}                  & \begin{tabular}[c]{@{}c@{}}32.7\%\\ (28.5\%)\end{tabular}                       & +14.2\%                                                                                        \\ \hline
1.3                                                                                        & 2048                                                                            & 24                               & 16                                                                                  & 8                              & 8                                                                                 & 1                                                                                            & 1                                                                                   & 1                                                                                     & 64                                                                             & \multicolumn{1}{c|}{\begin{tabular}[c]{@{}c@{}}115,682\\ (109,537)\end{tabular}}              & \multicolumn{1}{c|}{\begin{tabular}[c]{@{}c@{}}39,775\\ (39,537)\end{tabular}}            & \multicolumn{1}{c|}{\begin{tabular}[c]{@{}c@{}}150\\ (142)\end{tabular}}                 & \begin{tabular}[c]{@{}c@{}}48.1\%\\ (45.5\%)\end{tabular}                       & +5.6\%                                                                                         \\ \hline
6.7                                                                                        & 4096                                                                            & 32                               & 32                                                                                  & 16                             & 1                                                                                 & 16                                                                                           & 1                                                                                   & 1                                                                                     & 128                                                                            & \multicolumn{1}{c|}{\begin{tabular}[c]{@{}c@{}}44,605\\ (39,936)\end{tabular}}                & \multicolumn{1}{c|}{\begin{tabular}[c]{@{}c@{}}35,442\\ (35,403)\end{tabular}}            & \multicolumn{1}{c|}{\begin{tabular}[c]{@{}c@{}}149\\ (116)\end{tabular}}                 & \begin{tabular}[c]{@{}c@{}}47.8\%\\ (43.0\%)\end{tabular}                       & +11.7\%                                                                                        \\ \hline
175                                                                                        & 12288                                                                           & 96                               & 96                                                                                  & 128                            & 1                                                                                 & 1                                                                                            & 8                                                                                   & 16                                                                                    & 1536                                                                           & \multicolumn{1}{c|}{\begin{tabular}[c]{@{}c@{}}14,634\\ (14,571)\end{tabular}}                & \multicolumn{1}{c|}{\begin{tabular}[c]{@{}c@{}}34,912\\ (34,708)\end{tabular}}            & \multicolumn{1}{c|}{\begin{tabular}[c]{@{}c@{}}160\\ (159)\end{tabular}}                 & \begin{tabular}[c]{@{}c@{}}51.3\%\\ (50.9\%)\end{tabular}                       & +0.4\%                                                                                         \\ \hline
\end{tabular}
  }
  \label{fig:platform}%
\end{table*}%


\begin{table*}
  \centering
  \caption{Results for large-scale MoE training on GPT models with different configurations.}
  \vspace{-3mm}
  \resizebox{0.9\textwidth}{!}{%
  \begin{tabular}{|c|c|c|c|c|c|c|c|cc|cc|}
    \hline
    \multirow{2}{*}{\textbf{Parameters(B)}} &
      \multirow{2}{*}{\textbf{\begin{tabular}[c]{@{}c@{}}Attention\\ heads\end{tabular}}} &
      \multirow{2}{*}{\textbf{\begin{tabular}[c]{@{}c@{}}Hidden\\ size\end{tabular}}} &
      \multirow{2}{*}{\textbf{\begin{tabular}[c]{@{}c@{}}Vocab\\ size\end{tabular}}} &
      \multirow{2}{*}{\textbf{Layers}} &
      \multirow{2}{*}{\textbf{Experts}} &
      \multirow{2}{*}{\textbf{GPUs}} &
      \multirow{2}{*}{\textbf{\begin{tabular}[c]{@{}c@{}}Batch\\ size\end{tabular}}} &
      \multicolumn{2}{c|}{\textbf{Speed(tokens/s)}} &
      \multicolumn{2}{c|}{\textbf{Memory(GB)}} \\ \cline{9-12} 
     &
       &
       &
       &
       &
       &
       &
       &
      \multicolumn{1}{l|}{DeepSpeed} &
      \multicolumn{1}{l|}{\TheName{}} &
      \multicolumn{1}{l|}{DeepSpeed} &
      \multicolumn{1}{l|}{\TheName{}} \\ \hline
    13.9 &
      \multirow{5}{*}{64} &
      \multirow{5}{*}{4096} &
      \multirow{5}{*}{50304} &
      \multirow{5}{*}{12} &
      8 &
      8 &
      8 &
      \multicolumn{1}{c|}{24165} &
      \textbf{31085} &
      \multicolumn{1}{c|}{68.9} &
      \textbf{56.8} \\ \cline{1-1} \cline{6-12} 
    26.8 &
       &
       &
       &
       &
      16 &
      16 &
      16 &
      \multicolumn{1}{c|}{43691} &
      \textbf{59136} &
      \multicolumn{1}{c|}{66.2} &
      \textbf{53.9} \\ \cline{1-1} \cline{6-12} 
    52.6 &
       &
       &
       &
       &
      32 &
      32 &
      32 &
      \multicolumn{1}{c|}{82957} &
      \textbf{113456} &
      \multicolumn{1}{c|}{66.8} &
      \textbf{54.5} \\ \cline{1-1} \cline{6-12} 
    104.1 &
       &
       &
       &
       &
      64 &
      64 &
      64 &
      \multicolumn{1}{c|}{157728} &
      \textbf{209970} &
      \multicolumn{1}{c|}{66.3} &
      \textbf{54.4} \\ \cline{1-1} \cline{6-12} 
    207.2 &
       &
       &
       &
       &
      128 &
      128 &
      128 &
      \multicolumn{1}{c|}{283706} &
      \textbf{376968} &
      \multicolumn{1}{c|}{66.4} &
      \textbf{54.3} \\ \hline
    \end{tabular}
  }
  \label{tab:exp_training}%
\end{table*}%

\begin{table*}
  \centering
  \caption{Results for elastic MoE training with multiple tasks.}
  \vspace{-3mm}
  \resizebox{0.9\textwidth}{!}{%
  \begin{tabular}{|c|c|c|c|c|c|c|c|c|}
  \hline
   &
    \textbf{Task number} &
    \textbf{Parameters(M)} &
    \textbf{\begin{tabular}[c]{@{}c@{}}Total\\ batch size\end{tabular}} &
    \textbf{\begin{tabular}[c]{@{}c@{}}Batch size \\ per task\end{tabular}} &
    \textbf{GPUs} &
    \textbf{\begin{tabular}[c]{@{}c@{}}GPUs\\ per task\end{tabular}} &
    \textbf{\begin{tabular}[c]{@{}c@{}}Total Speed\\ (samples/s)\end{tabular}} &
    \textbf{\begin{tabular}[c]{@{}c@{}}Speed per GPU\\ (samples/s)\end{tabular}} \\ \hline
  Load imbalanced &
    \multirow{2}{*}{4} &
    \multirow{2}{*}{83} &
    \multirow{2}{*}{1024} &
    \multirow{2}{*}{512/256/128/128} &
    4 &
    1/1/1/1 &
    250.4 &
    62.6 \\ \cline{1-1} \cline{6-9} 
  Load balanced &
     &
     &
     &
     &
    8 &
    4/2/1/1 &
    591.9 &
    74.0 \\ \hline
  \end{tabular}%
  }
  \label{tab:exp_load_balance}%
\end{table*}

\subsection{Large-Scale MoE Training}

We conducted training experiments using GPT models based on the MoE architecture, leveraging A100 GPUs (80 GB), and employing a combination of data parallelism and expert parallelism. In the MoE system, there are two main parts: the Dense parameters (Backbone) and the Sparse parameters (Expert). The Dense part employs data parallelism, meaning that different input data is processed in parallel. After the backward computation is complete, the Dense parameters are synchronized through Allreduce communication. On the other hand, the Sparse parameter part involves expert parallelism. Here, routing communication between experts is used to send the required data to the designated compute devices. This is accomplished through AlltoAll communication. The backward pass also utilizes AlltoAll communication for gradient synchronization. This approach allows us to efficiently leverage both data and expert parallelism, optimizing the performance of the MoE system. The evaluation of these models was performed using Gshard~\cite{gross2017hard} and top1-gating metrics. Specifically, we utilized pure fp16 precision and the AdamW~\cite{loshchilov2018decoupled} optimizer during the training process.

Compared to the high precision FP32 method, there are two lower precision training approaches: "AMP" (Automatic Mixed Precision) and "pure fp16". Unlike AMP, pure fp16 is a commonly used, faster training method. This method is well-established and has been applied in model training, for example in Megatron and DeepSpeed. Our approach is consistent with the methods used in Megatron/DeepSpeed. The term "pure" in "pure fp16" is in contrast to AMP, meaning that all model parameters in pure fp16 training are of fp16 type. However, it's important to note that not all operations are computed in fp16. Certain operations, like softmax, use fp32 for computation. We also employ techniques like MasterWeight to mitigate the impact of lower precision training on model accuracy. Our use of pure fp16 in the experiments is a standard practice, and we maintain this strategy consistently when comparing with other frameworks.

Table~\ref{tab:exp_training} presents the throughput results for different configurations. The table lists the parameter sizes (\emph{Parameters (B)} in billions) in ascending order from top to bottom, while keeping the number of attention heads (\emph{Attention heads}), hidden layer size (\emph{Hidden size}), vocabulary size (\emph{Vocab size}), and number of layers (\emph{Layers}) constant. Consequently, the number of experts (\emph{Experts}), GPUs (\emph{GPUs}), and batch size (\emph{Batch size}) increase twofold. Both DeepSpeed and our proposed \TheName{} exhibit training speeds that double accordingly. It is worth noting that the first line of the table represents the performance on a single node equipped with eight GPUs, while the subsequent lines depict results for multi-node scenarios.

Our observations indicate that, compared to the state-of-the-art MoE system, DeepSpeed~\footnote{\url{https://github.com/microsoft/Megatron-DeepSpeed}}, \TheName{} achieves approximately a 28\% speedup in single-node training and at least a 33\% speedup in multi-node training for MoE models with over 100 billion parameters. Furthermore, \TheName{} reduces the GPU memory usage of each rank by nearly 12 GB. Therefore, in large-scale MoE training, our proposed \TheName{} system demonstrates comparable training speeds while consuming relatively less memory compared to the benchmark model, DeepSpeed.

\subsection{Ablation Study in MoE Training}
\label{exp:ela_train}


The experimental evaluation in this section highlights the benefits of efficient \textbf{\emph{implementation strategies}} on a large-scale model, as discussed in Section~\ref{sec:eff_methods}. Each of these strategies is evaluated independently and compared against traditional/baseline methods.

\subsubsection{Elastic MoE Training}

In order to assess the efficiency of elastic MoE training, we conducted experiments using the UFO~\cite{zhang2021federated} model, which is based on the MoE architecture and trained on A100 GPUs with 80 GB of memory. We designed four tasks with batch sizes of 512, 256, 128, and 128, respectively, to simulate an imbalanced training process.

Following the elastic sparse training methodology outlined in Section~\ref{sec:elastic_training}, we adjusted the overall training workload by adding additional computing nodes. Specifically, we allocated 4 GPUs for Task-1 and 2 GPUs for Task-2. To ensure fairness, we calculated the average speed of each GPU to mitigate the impact of increasing the number of nodes. The results, as presented in Table~\ref{tab:exp_load_balance}, indicate that compared to the \emph{Load imbalanced} configuration, the \emph{Load balanced} configuration achieved an approximate 18.2\% improvement in throughput per GPU. It is important to note that the \emph{Load balanced} approach was derived from our designed elastic MoE training, and the results highlight its effectiveness and efficiency, respectively.

Furthermore, we applied a task-based MoE load balancing mechanism to the training of the billion-scale visual model VIMER-UFO 2.0. This approach supports dynamic expansion of task numbers and parallel training of multiple tasks and multiple experts. Under the same experimental environment (32x A100 80GB GPUs), we achieved a training performance of 697 images per second. This represents a significant improvement in throughput by 64\% compared to 425 images per second using the Pytorch v1.10 framework. Additionally, the memory footprint was reduced to 45 GB per GPU, a decrease of 18\%.

\subsubsection{Resource-Aware Communication}

In this subsection, we conducted training of MoE models on different numbers of nodes and with varying model sizes to demonstrate the potential benefits of the \TheName{} designs. The results are presented using a stacked bar chart, which illustrates the time consumption for key components of the training process: 1) the forward stage (\emph{FWD}), 2) the backward stage (\emph{BWD}), 3) the optimization stage (\emph{OPT}), and 4) the communication stage. Specifically, we compare the proposed \emph{Hierarchical AlltoAll} approach to the baseline \emph{AlltoAll} method in the communication stage, while keeping other components constant across the different settings.

\begin{figure}[h!]
  \centering
  \subfloat[2 nodes with 16 GPUs]{\includegraphics[width=0.48\columnwidth]{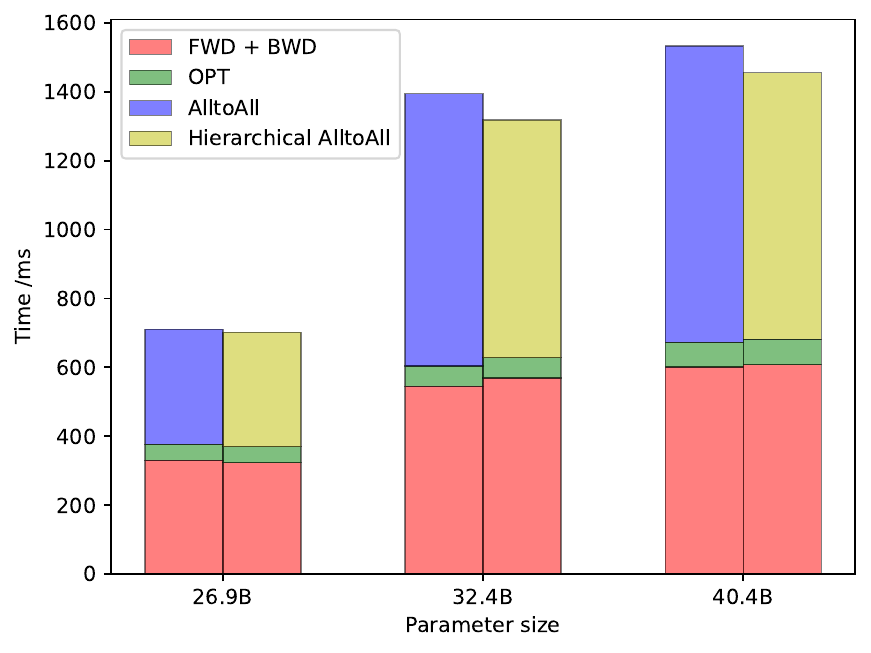}
  \label{fig:adaptive_original}}
    \hfil
  \subfloat[4 nodes with 32 GPUs]{\includegraphics[width=0.48\columnwidth]{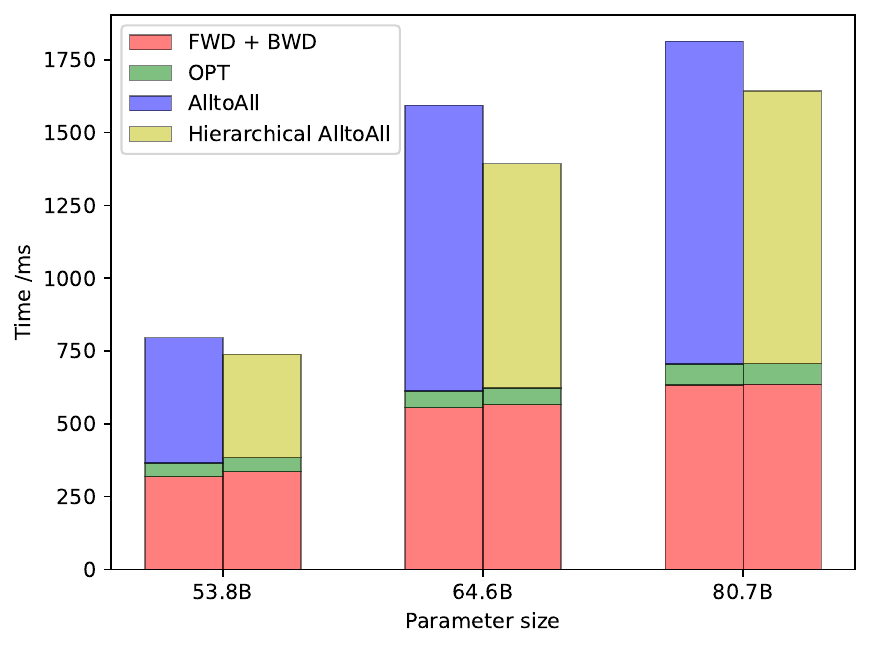}
  \label{fig:adaptive_fault}}
 \vspace{-10pt}
  \caption{MoE Training Time Breakdown}
  \label{fig:exp_hierarchical_alltoall}
\end{figure}

\begin{table*}
  \centering
  \caption{Performance of the Embedding Partition in Data Parallelism.}
 \vspace{-3mm}
  \resizebox{0.9\textwidth}{!}{%
  \begin{tabular}{|c|c|c|c|c|c|cc|cc|}
  \hline
  \multirow{2}{*}{\textbf{\begin{tabular}[c]{@{}c@{}}Batch\\ size\end{tabular}}} & \multirow{2}{*}{\textbf{GPUs}} & \multirow{2}{*}{\textbf{Experts}} & \multirow{2}{*}{\textbf{\begin{tabular}[c]{@{}c@{}}Vocab\\ size\end{tabular}}} & \multirow{2}{*}{\textbf{\begin{tabular}[c]{@{}c@{}}Hidden\\ size\end{tabular}}} & \multirow{2}{*}{\textbf{Parameter(M)}} & \multicolumn{2}{c|}{\textbf{\begin{tabular}[c]{@{}c@{}}Memory (GB)\end{tabular}}}  & \multicolumn{2}{c|}{\textbf{\begin{tabular}[c]{@{}c@{}}Speed (tokens/s)\end{tabular}}} \\ 
    \cline{7-10} 
        & & & & & & \multicolumn{1}{l|}{\textbf{Baseline}} & \multicolumn{1}{l|}{\textbf{Embedding Partition}} & \multicolumn{1}{l|}{\textbf{Baseline}} & \multicolumn{1}{l|}{\textbf{Embedding Partition}} \\ 
        \hline
  \multirow{3}{*}{8} & \multirow{3}{*}{8} & \multirow{3}{*}{\textbf{4}} & \multirow{3}{*}{50304} & 2048 & 72 & \multicolumn{1}{c|}{4.68} & \textbf{2.43} & \multicolumn{1}{c|}{289452} & \textbf{300451} \\ 
    \cline{5-10} 
  & & & & 4096 & 180 & \multicolumn{1}{c|}{7.51} & \textbf{4.67} & \multicolumn{1}{c|}{152144} & \textbf{167352} \\ \cline{5-10} 
        & & & & 8192 & 400 & \multicolumn{1}{c|}{15.81} & \textbf{8.63} & \multicolumn{1}{c|}{80421} & \textbf{91687} \\ 
        \hline

        \multirow{3}{*}{8} & \multirow{3}{*}{8} & \multirow{3}{*}{\textbf{8}} & \multirow{3}{*}{50304} & 2048 & 100 & \multicolumn{1}{c|}{7.46} & \textbf{5.78} & \multicolumn{1}{c|}{144159} & \textbf{150161} \\ 
    \cline{5-10} 
  & & & & 4096 & 300 & \multicolumn{1}{c|}{12.80} & \textbf{9.70} & \multicolumn{1}{c|}{86237} & \textbf{95890} \\ \cline{5-10} 
        & & & & 8192 & 700 & \multicolumn{1}{c|}{27.80} & \textbf{20.49} & \multicolumn{1}{c|}{40605} & \textbf{46938} \\ 
        \hline

        \multirow{3}{*}{8} & \multirow{3}{*}{8} & \multirow{3}{*}{\textbf{16}} & \multirow{3}{*}{50304} & 2048 & 227 & \multicolumn{1}{c|}{18.55} & \textbf{15.73} & \multicolumn{1}{c|}{53428} & \textbf{55725} \\ 
    \cline{5-10} 
  & & & & 4096 & 781 & \multicolumn{1}{c|}{31.41} & \textbf{27.98} & \multicolumn{1}{c|}{26321} & \textbf{29047} \\ \cline{5-10} 
        & & & & 8192 & 1320 & \multicolumn{1}{c|}{63.25} & \textbf{55.75} & \multicolumn{1}{c|}{12065} & \textbf{13902} \\ 
        \hline
  \end{tabular}%
  }
  \label{tab:embedding_partition}
\end{table*}

\begin{table*}
\centering
\caption{Performance comparison among the proposed strategies.}
\vspace{-3mm}
\resizebox{0.9\textwidth}{!}{%
\begin{tabular}{|c|c|l|c|c|cccc|cccc|}
\hline
\multirow{2}{*}{\textbf{\begin{tabular}[c]{@{}c@{}}Batch\\ size\end{tabular}}} &
  \multirow{2}{*}{\textbf{\begin{tabular}[c]{@{}c@{}}Vocab\\ size\end{tabular}}} &
  \multirow{2}{*}{\textbf{Layers}} &
  \multirow{2}{*}{\textbf{\begin{tabular}[c]{@{}c@{}}Hidden\\ size\end{tabular}}} &
  \multirow{2}{*}{\textbf{Experts}} &
  \multicolumn{4}{c|}{\textbf{\begin{tabular}[c]{@{}c@{}}Peak Memory (GB)\end{tabular}}} &
  \multicolumn{4}{c|}{\textbf{\begin{tabular}[c]{@{}c@{}}Speed per GPU (tokens/s)\end{tabular}}} \\ \cline{6-13} 
 &
   &
   &
   &
   &
  \multicolumn{1}{c|}{Baseline} &
  \multicolumn{1}{c|}{\begin{tabular}[c]{@{}c@{}}Elastic \\ Training\end{tabular}} &
  \multicolumn{1}{c|}{\begin{tabular}[c]{@{}c@{}}Resource-Aware \\ Communication\end{tabular}} &
  \begin{tabular}[c]{@{}c@{}}Embedding \\ Partition\end{tabular} &
  \multicolumn{1}{c|}{Baseline} &
  \multicolumn{1}{c|}{\begin{tabular}[c]{@{}c@{}}Elastic \\ Training\end{tabular}} &
  \multicolumn{1}{c|}{\begin{tabular}[c]{@{}c@{}}Resource-Aware \\ Communication\end{tabular}} &
  \begin{tabular}[c]{@{}c@{}}Embedding \\ Partition\end{tabular} \\ \hline
\multirow{2}{*}{8} &
  \multirow{2}{*}{50304} &
  \multicolumn{1}{c|}{\multirow{2}{*}{12}} &
  \multirow{2}{*}{2048} &
  \multirow{2}{*}{8} &
  \multicolumn{1}{c|}{\multirow{2}{*}{28}} &
  \multicolumn{1}{c|}{\multirow{2}{*}{23}} &
  \multicolumn{1}{c|}{\multirow{2}{*}{28}} &
  \multirow{2}{*}{25} &
  \multicolumn{1}{c|}{\multirow{2}{*}{30604}} &
  \multicolumn{1}{c|}{\multirow{2}{*}{35194}} &
  \multicolumn{1}{c|}{\multirow{2}{*}{33664}} &
  \multirow{2}{*}{31033} \\
 &
   &
  \multicolumn{1}{c|}{} &
   &
   &
  \multicolumn{1}{c|}{} &
  \multicolumn{1}{c|}{} &
  \multicolumn{1}{c|}{} &
   &
  \multicolumn{1}{c|}{} &
  \multicolumn{1}{c|}{} &
  \multicolumn{1}{c|}{} &
   \\ \hline
\end{tabular}%
}
\label{tab:cross-wise}
\end{table*}

As depicted in Figure~\ref{fig:exp_hierarchical_alltoall}, the first three components (green and pink bars) show similar performance between the AlltoAll baseline and Hierarchical AlltoAll across all parameter sizes. The performance gaps are primarily caused by the two top bars (purple for AlltoAll baseline and yellow for Hierarchical AlltoAll) representing the communication stage. It is evident that with the adoption of Hierarchical AlltoAll in \TheName{}, the computation time does not increase significantly, while the communication time decreases dramatically. Furthermore, as the model parameter size increases, the efficiency gap in communication becomes more significant between AlltoAll and hierarchical AlltoAll. This indicates that the proposed \TheName{} can amplify the performance improvement in communication when dealing with large-scale models. This effect is evident by observing the gap between the purple and yellow bars from left to right along the horizontal axis.

For the most substantial improvement observed in the experiment, which involved a MoE model with 80.7 billion parameters across four nodes with 32 GPUs, the overall end-to-end training performance improved by 10.3\%. Additionally, the communication stage achieved a 15.5\% speedup using the Hierarchical AlltoAll strategy.

\subsubsection{Embedding Partition in Data Parallelism} 
To assess the performance of embedding partition in data parallelism, we conducted training of a MoE model on a dataset with an extremely large vocabulary. The ablation study included a baseline approach using the non-segment embedding strategy, while the remaining settings remained consistent with \TheName{}. The results in Table~\ref{tab:embedding_partition} demonstrate that the embedding partition strategy within a single machine effectively reduces GPU memory consumption when processing vocabularies of large sizes. The study maintains a constant batch size and number of GPUs but varies the number of experts, hidden size, and parameter magnitude. Embedding partitioning consistently exhibits significant reductions in memory usage (e.g., a drop from 15.81 GB to 8.63 GB with 4 experts and hidden size 8192) and enhancements in processing speed (as seen with the increase from 80421 to 91687 tokens/s for the same configuration). These improvements are maintained across various model complexities, indicated by varying numbers of parameters and experts, suggesting that embedding partitioning provides a scalable solution for optimizing large-scale neural networks in data-parallel scenarios.  

\subsubsection{Cross-wise Comparison}


In order to determine the dominant strategy and its contribution to the overall performance gain in the \TheName{} system, we conducted a cross-wise comparison among the proposed strategies. While we have individually showcased the performance gains of each strategy compared to the baselines, their relative performance among each other requires further investigation. We are particularly interested in identifying the strategy that is most dominant and contributes the most to the \TheName{} system. To conduct this comparison, we performed experiments measuring the peak memory usage and average computation speed on the GPU in parallel. The results are summarized in Table~\ref{tab:cross-wise}. The baseline refers to the MoE without any of the proposed strategies, and as expected, it exhibits the highest peak memory consumption and the lowest computation speed. Among the proposed strategies, Elastic Training shows the least peak memory occupancy, while the Hierarchical AlltoAll strategy (utilized for resource-aware communication) consumes relatively more peak memory. Regarding GPU computation speed, Elastic Training again outperforms the other three strategies.

As shown in Fig.~\ref{fig:cross_all} directly, compared to a baseline scenario with no optimizations, our elastic training strategy shows the most significant improvements in terms of peak memory usage and GPU computation speed, contributing greatly to overall performance enhancement. Additionally, the topology-aware hierarchical AllToAll strategy and the DP-Embedding sharding strategy also contribute to a noticeable proportion of performance improvement in MoESys.

\subsection{MoE Inference}
\label{sec:infer_exp}


The inference experiments consist of two parts: the first part evaluates the performance of the MoE inference system using different models with varying numbers of parameters (in the billions), while the second part assesses the effectiveness of the offloading strategy proposed in Section~\ref{sec:ring_memory}.

\subsubsection{Effectiveness}

\begin{table}[h!]
  \centering
  \caption{Performance of MoE inference.}
 \vspace{-3mm}
  \resizebox{0.4\textwidth}{!}{
    \begin{tabular}{|c|c|c|cc|}
    \hline
    \multirow{2}{*}{\textbf{Parameters(B)}} &
      \multirow{2}{*}{\textbf{GPUs}} &
      \multirow{2}{*}{\textbf{\begin{tabular}[c]{@{}c@{}}Batch\\ size\end{tabular}}} &
      \multicolumn{2}{c|}{\textbf{Speed(tokens/s)}} \\ \cline{4-5} 
        &    &    & \multicolumn{1}{l|}{DeepSpeed} & \multicolumn{1}{l|}{\TheName{}} \\ \hline
    10.0  & 1  & 1  & \multicolumn{1}{c|}{4303}      & \textbf{4551}               \\ \hline
    106.5 & 8  & 8  & \multicolumn{1}{c|}{27215}     & \textbf{29681}              \\ \hline
    209.6 & 16 & 16 & \multicolumn{1}{c|}{35310}     & \textbf{40059}              \\ \hline
    \end{tabular}
  }
  \label{tab:exp_inference}%
\end{table}%

Inference tasks typically require less memory compared to training tasks in many scenarios. Consequently, it is feasible to perform downstream tasks using a single GPU with a 10-billion-parameter MoE model. To evaluate the inference performance of large-scale MoE models, we conducted experiments on a text generation task. The results presented in Table~\ref{tab:exp_inference} indicate that \TheName{} achieves approximately 13\% faster inference speed compared to DeepSpeed for MoE models with over 200 billion parameters. This considerable performance improvement underscores the effectiveness of \TheName{} in the inference process.

\subsubsection{Ring Memory Offloading}

\begin{figure}[htbp]
\vspace{-10pt}
\centering
\begin{minipage}{.23\textwidth}
  \centering
  \includegraphics[width=.98\linewidth]{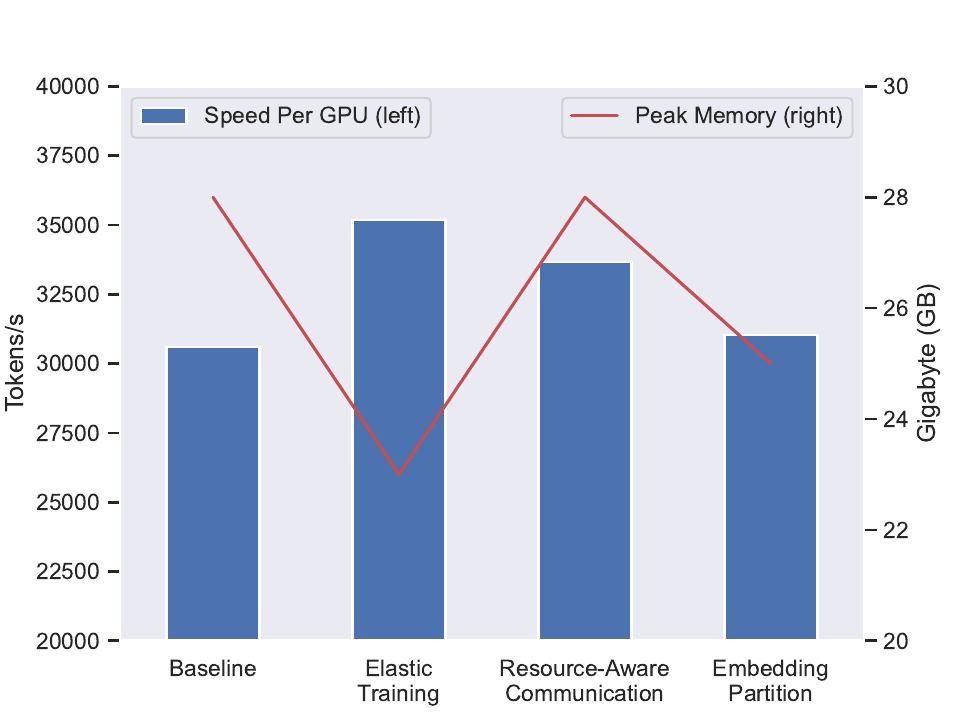}
  \vspace{-10pt}
  \captionof{figure}{Cross-wise comparison among proposed strategies.}
  \label{fig:cross_all}
\end{minipage}%
\hspace{8pt}
\begin{minipage}{.22\textwidth} 
  \centering
  \includegraphics[width=.99\linewidth]{figures/comparison_all.pdf}
  \vspace{-10pt}
  \captionof{figure}{Evaluation of MoE inference with and without overlapping offloading.}
  \label{fig:exp_ring_memory}
\end{minipage}
\vspace{-10pt}
\end{figure}

We conducted experiments to evaluate the inference performance of the expert offloading strategy using ring memory on a system equipped with 16 A100(40G) GPUs. The experiment focused on a MoE model with 32 experts and 58.2 billion parameters. We measured the time consumed for computation in GPU memory as well as the movement of experts between CPU and GPU memory. Figure~\ref{fig:exp_ring_memory} illustrates the results, indicating that the performance of the overlapped MoE inference system remains largely unaffected by CPU offloading. The findings demonstrate that this strategy achieves a favorable balance between computation and data movement. Additionally, it enables the MoE inference systems to reduce GPU memory consumption by at least 30\% compared to inference without ring memory offloading.


\subsection{Summary}

The aforementioned experiments comprehensively examine the efficiency and efficacy of the proposed designs within the \TheName{} systems. Particularly, when applied to large-scale deep learning models such as the GPT series, \TheName{} exhibits superior performance in terms of training and inference speed, as well as memory consumption, in comparison to the established benchmark DeepSpeed. As an industrially-relevant MoE system, \TheName{} is demonstrated to further enhance the development of distributed MoE designs in practical real-world applications.

\section{Conclusion and Future Works}

To address the needs of modern internet services that demand the use of large-scale DNNs, we have presented \TheName{}, a novel training and inference system based on MoE that boosts efficiency in both large-scale training and inference. Our proposed system adopts an elastic training strategy with 2D prefetch and fusion communication over hierarchical storage for efficient parallelisms. For scalable inference in a single node, \TheName{} builds CPU-GPU memory jointly into a ring of sections and executes computation tasks across memory sections in a round-robin manner. Our experiments demonstrate that \TheName{} achieves superior performance compared to the state-of-the-art DeepSpeed, outperforming DeepSpeed by 33\% in training throughput and 13\% in inference throughput, with 64\% higher throughput and 18\% lower memory footprints under MoE tasks based on large language models and foundation vision models. 

Our future work focuses on developing a unified sparse training and inference system that considers parameter-server and scheduling across multiple dimensions, exploring efficient methods for sparse training within the \TheName{} framework, and enhancing our system collaboration with resource platforms for sustainable research. We also aim to implement a comprehensive evaluation framework that accurately assesses the comparative performance of diverse parallel computing strategies in a way that is fair, informative, and contributes constructively to the field of scalable machine learning architectures.


\bibliographystyle{IEEEtran}
\bibliography{main}

\end{document}